\documentclass[prl,twocolumn,showpacs,superscriptaddress,amsmath,amssymb]{revtex4}
\usepackage{graphicx}

\newcommand{\ghh}{\ensuremath{\Gamma_{8,\text{HH}}}}
\newcommand{\glh}{\ensuremath{\Gamma_{8,\text{LH}}}}
\newcommand{\gs}{\ensuremath{\Gamma_{6}}}

\begin{document}
\title{Topological surface states of strained Mercury-Telluride probed by ARPES}
\author{Olivier Crauste}
\affiliation{Institut N\'eel, C.N.R.S. and Universit\'e Joseph Fourier, BP 166,
38042 Grenoble Cedex 9, France}
\author{Yoshiyuki Ohtsubo}
\affiliation{Synchrotron SOLEIL, Saint-Aubin BP 48, F-91192 Gif-sur-Yvette, France}
\author{Philippe Ballet}
\affiliation{CEA, LETI, MINATEC Campus, DOPT, 17 rue des martyrs 38054 Grenoble Cedex 9, France}
\author{Pierre Delplace}
\affiliation{D\'epartement de Physique Th\'eorique, Universit\'e de Gen{\`e}ve, CH-1211 Gen{\`e}ve 4, Switzerland}
\author{David Carpentier}
\affiliation{Laboratoire de Physique, Ecole Normale Sup{\'e}rieure de Lyon and CNRS
UMR5672, France}
\author{Cl\'{e}ment Bouvier}
\affiliation{Institut N\'eel, C.N.R.S. and Universit\'e Joseph Fourier, BP 166,
38042 Grenoble Cedex 9, France}
\author{Tristan Meunier}
\affiliation{Institut N\'eel, C.N.R.S. and Universit\'e Joseph Fourier, BP 166,
38042 Grenoble Cedex 9, France}
\author{Amina Taleb-Ibrahimi}
\affiliation{Synchrotron SOLEIL, Saint-Aubin BP 48, F-91192 Gif-sur-Yvette, France}
\author{Laurent P. L\'{e}vy}
\affiliation{Institut N\'eel, C.N.R.S. and Universit\'e Joseph Fourier, BP 166,
38042 Grenoble Cedex 9, France}
\email[Corresponding author:]{Laurent.Levy@grenoble.cnrs.fr}
\homepage{http://www.neel.cnrs.fr/}

\date{\today}

\begin{abstract}
The topological surface states of strained HgTe have been measured using high-resolution ARPES measurements.  The dispersion of surface states form a Dirac cone, which origin is close to the top of the \ghh band: the top half of the Dirac cone is inside the stress-gap while the bottom half lies within the heavy hole bands and keeps a linear dispersion all the way to the X-point. The circular dichroism of the photo-emitted electron intensity has also been measured for all the bands.
\end{abstract}


\maketitle{}

Topological Insulators are new states of matter possessing a topological order induced by a strong spin-orbit.  One of the hallmark of this topological order is the appearance of robust surface states, whose precise nature is a unique signature of the unconventional bulk order.  In the case of the three dimensional topological insulators, these surface states consist in an odd number of species of Dirac particles, as opposed to 2D systems where Dirac states can only appear in pairs. Moreover, these surface Dirac states display remarkable magnetic textures, reflecting their spin-orbit origin where momentum and spin are bound together.  While transport experiments proved to be poor probes of surface states, Angle Resolved PhotoEmission Spectroscopy (ARPES) appeared as
a tool of choice. Landmark ARPES experiments on Bismuth-Antimony\cite{Hsieh2008,Hsieh2009,Hsieh2010}, Bismuth-Telluride\cite{Chen2009,Hajlaoui2012}, Bismuth-Selenide\cite{Analytis2010, Kuroda2010,King2011,Wang2011,Park2012} have firmly established topological insulators as real materials.  A number of other experimental tools, such as surface STM-AFM studies\cite{Hanaguri2010}, magnetotransport\cite{Analytis2010}, orbital magnetometry\cite{Taskin2009}, far-infrared spectroscopy\cite{Orlita2011}, have revealed a number of the physical properties of the surface states of these materials.  However, the cleanest evidences for topological surface states seem always to come from low energy ARPES. In this letter, we analyze the nature of surface states of strained Bulk Mercury Telluride in the ARPES spectra, and prove its topological structure.

Bulk Mercury Telluride is a semi-metal with an inverted band structure: the $\Gamma_8$ band is lying 0.3 eV above the $\Gamma_6$ band.  As pointed out in several seminal papers\cite{Bernevig2006,Fu2007}, this semi-metal can be turned into a two-dimensional spin-Hall insulator by confining the carriers inside a quantum well\cite{Konig2007}, or into a three-dimensional topological insulator by applying a bi-axial strain to the material\cite{Brune2011, Bouvier2012}. This can be achieved in the epitaxial growth of Mercury Telluride from a Cadmium Telluride substrate, which lattice constant is 0.3\% larger than Mercury Telluride. This strains opens a small gap (26 meV) between the $\Gamma_8$ light hole ($\glh$) and the $\Gamma_8$ heavy hole ($\ghh$) band.  Within this gap, the only conducting states left are topological surface states which give a metallic character to this three-dimensional topological insulator which bulk is truly insulating within the strain gap.

In this letter, the experimental ARPES spectra of homogenously strained 100 nm thick  HgTe slabs epitaxially grown on [100] CdTe substrates reveal very strong surface states whose Dirac point sits close to the top of the $\ghh$ band, leaving one half of the Dirac cone inside the strain gap and the other half inside the bulk $\ghh$ band. This near degeneracy between the Dirac point and the extremum of a bulk band is a unique feature found in no other topological insulating material.
These observations are confirmed by a discrete analysis of the 8-bands Kane model close to the $\Gamma$ point. Surprisingly enough, the surface states dispersion is linear all the way from the $\Gamma$ to the X point.  We have also measured the circular dichroism of the surface and bulk bands. We find a strong circular dichroism of the bulk \ghh{} band but no dichroism coming from the surface states. These new experimental findings are critical to the interpretation of transport experiments and to spintronic applications using strained Mercury Telluride.

\begin{figure*}[!t]
	\centering
	\includegraphics[width=\linewidth]{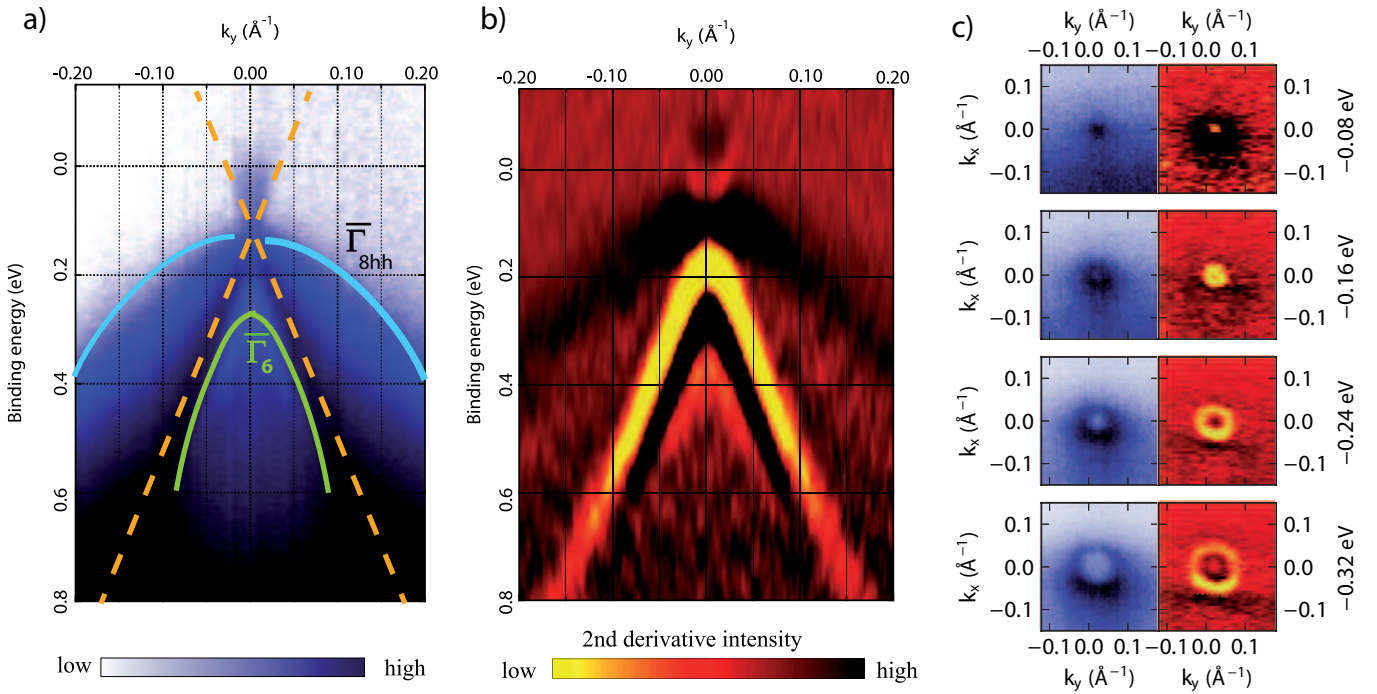}
	\caption{\label{fig:arpes_results}High resolution ARPES spectra for a maximally strained [100] HgTe/vacuum interface in the vicinity of the $\Gamma$-point measured at room temperature.
		a) Energy-momentum intensity spectrum after background substraction.
		b) The second derivative of the intensity data which band positions are less faithful enhances the contrast.
		c) Intensity spectrum at different energies. Raw data on the left and its second derivative on the right. The cone structure has a circular section up to $\approx 0.4$ eV.}
\end{figure*}

The ARPES-data were obtained on the CASSIOPEE line at the SOLEIL synchrotron\cite{URL-synchrotron}, which low energy photons and high resolution (few meV) spectrometer are well suited to topological insulator studies.  The strained HgTe slabs were grown by low temperature Molecular Beam Epitaxy in a Ultra-High Vacuum chamber from a [100] CdTe substrate which lattice constant ($a_{\text{CdTe}} = 6.48 \text{\AA{}}$) is 0.3\% larger than bulk HgTe ($a_{\text{HgTe}} = 6.46 \text{\AA{}}$).  As long as the HgTe thickness does not exceed $\approx 150\:$nm, the HgTe is expanded homogenously, as was checked by $\theta-2\theta$ X-ray scans and reciprocal space maps.  Only occupied electronic states are observed in ARPES: indium-doped samples at $10^{18} \text{cm}^{-3}$ were prepared in addition to un-doped reference samples, in order to raise the bulk chemical potential.

The samples surfaces were cleaned in a dedicated Ultra High Vacuum preparation chamber by a low-energy Ar-ion sputtering at grazing angles to remove the surface oxide. The
sharp dots observed in the \textsl{in-situ} LEED spectra showed that the surface was clean enough for the ARPES experiments\cite{surface-roughness}.  The samples were subsequently transferred to the ARPES chamber in Ultra-High Vacuum.  The position of the Fermi level was determined with a reference gold sample placed on the same sample holder.

We first present the high-resolution spectra in the vicinity of the $\Gamma$-point for an un-doped sample.  On the panel a) of Fig.~\ref{fig:arpes_results}, the intensity of the ARPES spectrum is shown for a incident photon energy $h\nu = 20\text{ eV}$. We retrieve the surface projection of the two volume valence bands \ghh{} and \gs{} (deep blue) and, with more intensity, a linear cone structure, which broadens as one moves away from its apex. The second derivative spectrum shown on panel b) enhances the contrast in the ARPES intensity. Within the experimental accuracy the cone apex coincide with the top of the \ghh{} band and lies 0.1 eV below the Fermi level.  On the raw ARPES spectrum shown in panel a) the cone structure extends in the gap with a decreasing intensity, as those states are populated mostly through the room-temperature thermal activation.  The cone section for different binding energies shown on the panel c) are circular up to energies 0.4 eV below the Dirac point.
From the experimental slope of the cone structure, the surface state band velocity is found to be $v_F \approx 5\times 10^{5} \text{m.s}^{-1}$.  This value agrees with the lowest order expansion for the energy close to the Dirac point in the Kane model ($\hbar v_F\approx\alpha\frac{P}{\sqrt{6}}$), where the parameter $\alpha\approx0.9$ for HgTe (the Kane parameters are defined in the supplementary material).
The same sample was also probed at different incident photon energies $h\nu$.  Varying the incident photon energy, shifts the binding energy of bulk bands according to their $k_z$ dispersion. Here, the cone position is unaffected, emphasizing that this cone structure comes from a surface state with no $k_z$ dispersion (see supplementary material, Fig.~1S). This is a powerful check which discriminates between 2D and 3D states.
 Surface state spectra were also collected over the entire Brillouin zone.  In the $\Gamma$-K direction, the surface state spectrum becomes diffuse at energies of 0.8 eV below the Fermi level.  On the other hand, in the $\Gamma$-X direction, the surface state spectra remain linear all the way the the X point (Supplementary material Fig.~2S), where its energy is 3.4 eV below the Dirac point, i.e. well below the \gs{} band: in this direction, the surface state robustness goes well beyond the usual topological protection arguments.
The ARPES spectra of doped samples are quite similar to the one presented in Fig. \ref{fig:arpes_results}, i.e. the electrochemical potential at the top surface appears to be little affected at the doping level used ($10^{18} \text{cm}^{-3}$-measured by an {\em ex-situ} SIMS analysis- which is equivalent to a surface density of $10^{13} \text{cm}^{-2}$ for a 100 nm thick slab).

\begin{figure}[!t]
	\centering
	\includegraphics[width=0.9\linewidth]{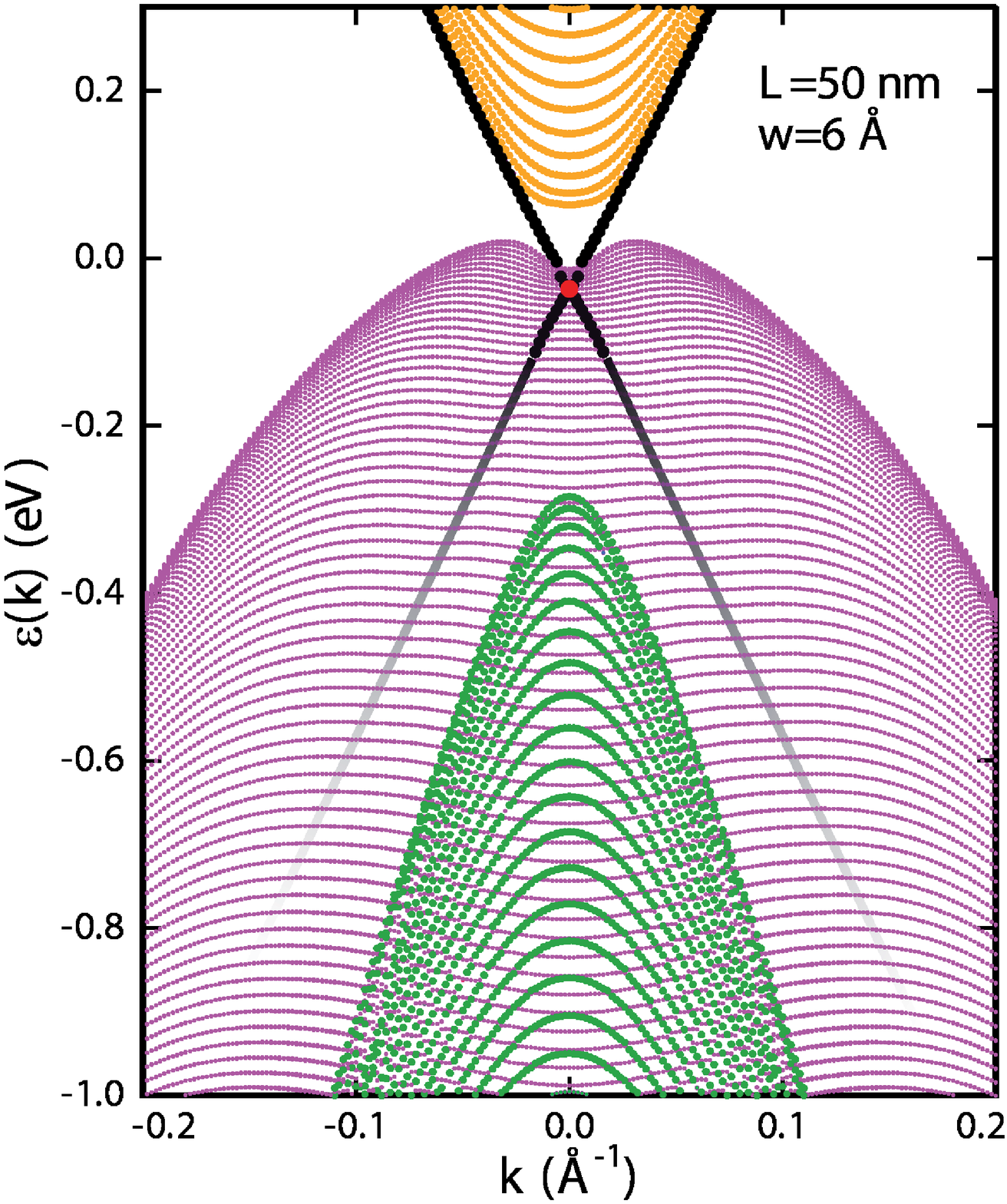}
	\caption{\label{fig:th}Surface (black) and bulk bands (\glh{} orange, \ghh{} magenta, \gs{} red) dispersions in the vicinity of the $\Gamma$ point computed with a discretized Kane model along the $z$ axis for a HgTe -- vacuum interface interface one lattice constant wide ($w$). The black dots are the computed surface states energies.}
\end{figure}

Previous theoretical studies\cite{Dai2008,Chu2011a} of HgTe surface states do not account quantitatively with our experimental findings in some of their salient features: weak hybridation between surface states and the \ghh{} valence band at small $k$ and a Dirac point which lies very close to the top of the \ghh{} valence band. Considering the successes of the Kane model\cite{Winkler2003}, for which all the parameters\cite{Novik2005} are known for HgTe, we computed the surface and bulk states of 2D interfaces between a maximally strained
 HgTe slab and vacuum (resp. CdTe) within the 8 bands ($\Gamma_{6,\pm 1/2}$, $\Gamma_{8, \pm 3/2, \pm 1/2}$ and $\Gamma_{7, \pm 1/2}$) Kane model.
The interfaces were described by interpolating smoothly over a finite width $w$
the Kane parameters between their values in HgTe for $0<z<L$ and vacuum (resp. CdTe) for $z<0$ and $z>L$. The corresponding Schr\"{o}dinger equation is discretized only in the $z$ direction and numerically solved, determining surfaces and bulk states \emph{at once}.  The results are carefully shown to be independent of the discretization constant $a$ over the range of energy and momenta considered. The results are shown in Fig.~\ref{fig:th} : the surface states, originating from the inversion between the two $S=\frac12$  bands $\Gamma_6$ and \glh{}, are the only states present in the gap. For the Kane parameters used, the energy of the Dirac point is $\epsilon_{D}=-30\:$meV below the \glh{} and is similar for a CdTe/HgTe interface.  At $k=0$ ($\epsilon_{D}$), the surface states do not couple to the \ghh{} band, and are weakly affected by the \ghh{} band at small $k$ below $\epsilon_{D}$. They disappear gradually for larger $k$, consistent with the observed broadening in the experiment. Their dispersion is linear with the same band velocity as in the experiment. As observed experimentally, half of the Dirac cone lies inside the \ghh{} valence band while the other half continues in the stress gap.

\begin{figure}
\centering
\includegraphics[width=\linewidth]{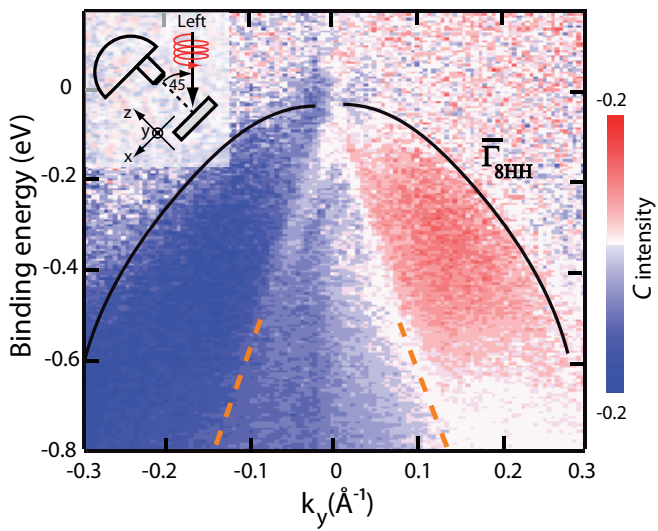}
\caption{\label{fig:dichroism}Circular dichroism measured at $k_x = 0${\AA}$^{-1}$. The largest contribution to the dichroism comes from the \ghh{} valence volume band. The surface states appear here as white lines (no dichroism).  The inset shows an incident left polarized light beam at $45^\circ$ with respect to the sample surface.}
\end{figure}

A hallmark of topological insulators is the helical spin structure of surface states induced by the strong spin-orbit coupling.  Such textures have been observed directly using spin-resolved ARPES.\cite{Hsieh2009,Hsieh2010,Kimura2010}  A less direct way to probe this helical spin-texture is through the induced circular dichroism in ARPES.\cite{Wang2011,Park2012}.   Circular dichroism is defined as the asymmetry between the ARPES intensity for left ($L$) and right ($R$) circular polarization
\begin{equation}
{\cal C}(\epsilon, k)=\frac{I_{R}(\epsilon, k) - I_{L}(\epsilon, k)}{I_{R}(\epsilon, k)+ I_{L}(\epsilon, k)}.
\end{equation}
It is plotted in Fig. \ref{fig:dichroism} (the geometry is specified in the inset) as a function of $k_y$ for $k_x = 0$ and an incident light beam at $\approx 45^\circ$ with respect to the normal to the sample.
By symmetry, the circular dichroism must cancel in the $k_y = 0$ plane as observed for incident photon energies $E_{\text{kin}} > 15.9\: \text{eV}$.

The most salient features of the experimental data of Fig. \ref{fig:dichroism} are (i) the absence of dichroism from the surface states, signaled by the  white lines (no dichroism) along the surface states dispersion and (ii) a significant dichroism (up to 20\%) is observed in the bulk \ghh{} band.  These results on strained HgTe differ from the circular dichroism ARPES data on Bi$_{2}$Se$_{3}$ compound \cite{Wang2011,Park2012} where a dominant signature of surface states was observed.
 The relationship between the circular dichroism and the spectral spin densities of low energy bands is complex and depends on the incident photon energies\cite{Scholz2013}. Hence, a quantitative description of circular dichroism may involve processes with higher energy bands which require ab-initio calculations for a proper treatment.
On the other hand if we assume that such a relationship exists,
Wang et al. \cite{Wang2011} have shown that the dependence of the ARPES polarization asymmetry on
the band polarizations, $\langle S_x \rangle$ and $\langle S_z \rangle$ is
\begin{equation}
 {\cal C}(\epsilon, k,\phi) =
-a^2 \cos{\phi}\langle S_z(\epsilon, k)\rangle + 4ab \sin\phi\langle S_x(\epsilon, k) \rangle,
\end{equation}
for a  circularly polarized light beam incident in the x-z plane at an angle $\phi$ with respect to the normal to the sample.  The matrix elements $a$ and $b$ depends on surface symmetries of the material \cite{Wang2011}.  This formula is consistent with the experimental data of Fig.\ref{fig:dichroism} if the coefficient $b$ vanishes for the [100] HgTe surface.
This explains the weak circular dichroism contribution of the surface states, whose spin polarization normal to the surface $\langle S_z(\epsilon, k)\rangle$ vanishes at low energy.
%
An interpretation along these lines also relates the observed dichroism of the valence band \ghh{} to its pseudo-spin polarization whose possible origins are well known\cite{Winkler2003}.  The first common source is the Rashba effect \cite{Bychkov1984,Winkler2003} : the $\text{Ar}^{+}$ beam used in the surface cleaning leaves a partially charged HgTe surface \cite{Solzbach1980}, inducing a surface electric field  $F$.
In addition, the Dresselhaus coupling present in centro-asymmetric crystals is here further enhanced by the strain deformation at the surface\cite{Ye2001}.  With these asymmetric couplings, the two \ghh{} spin-bands can acquire a sizable pseudospin polarization near the valence band maximum.
This spin structure has implications for the design of spintronic devices. In the stress gap, the only charge carriers are the helical surface states.  On the other hand, for negative energies the coexistence between a partially spin-polarized heavy-holes and helical surface states is less favorable for applications.

In summary, we have observed the surface states of stressed Mercury Telluride, confirming its topological insulating nature.  It has some quite unique features: the Dirac point sits at the top of the heavy hole band, something which was also noticed in transport experiment for samples oriented in the [100]\cite{Brune2011} and the [211]\cite{Bouvier2012} direction.  Combined with the very low residual bulk conduction in the stress gap, this topological insulator is one of the most interesting system for fundamental and applicative studies.

\begin{acknowledgments}
	The authors would like to thank H. Cercelier for his useful comments on the ARPES analysis. This work was funded by the EU contract GEOMDISS and the ANR grant SemiTopo.
\end{acknowledgments}

\bibliographystyle{apsrev}
\bibliography{arpes}

\begin{thebibliography}{26}
\bibitem{surface-roughness} The surface roughness was measured with an AFM on the oxydized sample surface after this experiment and found to be of the order of $\approx 2$ nm, i.e. $\approx$ three monolayers.
\end{thebibliography}
\end{document}